\documentclass[conference]{IEEEtran}
\IEEEoverridecommandlockouts
\usepackage{mathrsfs}
\usepackage{amsfonts}
\usepackage{ifpdf}
\usepackage{cite}
\usepackage{array}
\usepackage{booktabs}
\usepackage{setspace}

\ifCLASSINFOpdf
\usepackage[pdftex]{graphicx}
\else \fi
\usepackage[cmex10]{amsmath}
\usepackage{array}
\usepackage{mdwmath}
\usepackage{mdwtab}
\usepackage{eqparbox}
\usepackage[tight,footnotesize]{subfigure}
\usepackage{algorithmic}
\usepackage{algorithm}
\usepackage{amsmath}
\usepackage{epsfig}
\usepackage{ae}
\usepackage{amssymb}
\usepackage{times}
\usepackage{amsmath}   

\usepackage{xcolor}

\usepackage{amsthm} 

\usepackage{supertabular} 

\usepackage{multirow}
\newlength\savewidth

\newtheorem{theorem}{\textbf{Theorem}}
\newtheorem{lemma}{\textbf{Lemma}}

\newtheorem{remark}{\textbf{Remark}}

\begin{document}

\title{Sea Clutter Distribution Modeling: A Kernel Density Estimation Approach}
\author{\IEEEauthorblockN{Hongkuan Zhou, Yuzhou Li, and Tao Jiang}
\IEEEauthorblockA{School of Electronic Information and Communications, Huazhong University of Science and Technology, Wuhan, China \\
\{hongkuanzhou, yuzhouli, taojiang\}@hust.edu.cn
\thanks{This work was supported in part by the National Science Foundation of China with Grant Numbers 61601192, 61631015, and 61729101, the Young Elite Scientists Sponsorship Program by CAST with Grant Number 2017QNRC001, the State Key Laboratory of Integrated Services Networks (Xidian University) with Grant Number ISN19-09, and the Fundamental Research Funds for the Central Universities with Grant Numbers 2016YXMS298 and 2015ZDTD012.}
}
}
\maketitle

\IEEEpeerreviewmaketitle
\begin{abstract}
An accurate sea clutter distribution is crucial for decision region determination when detecting sea-surface floating targets.
However, traditional parametric models possibly have a considerable gap to the realistic distribution of sea clutters due to the volatile sea states. In this paper, we develop a kernel density estimation based framework to model the sea clutter distributions without requiring any prior knowledge.
In this framework, we jointly consider two embedded fundamental problems, the selection of a proper kernel density function
and the determination of its corresponding optimal bandwidth. Regarding these two problems, we adopt the Gaussian, Gamma, and Weibull distributions as the kernel functions, and derive the closed-form optimal bandwidth equations for them.
To deal with the highly complicated equations for the three kernels, we further design a fast iterative bandwidth selection
algorithm to solve them.
Experimental results show that, compared with existing methods, our proposed approach can significantly decrease the error incurred by sea clutter modeling (about two orders of magnitude reduction) and improve the target detection probability (up to $36\%$ in low false alarm rate cases).

\end{abstract}
\section{Introduction}
An important application for marine surveillance radar is to detect sea-surface small floating targets such as buoys, human divers, and small boats~\cite{SeaClutter_Iteration_TAES2017}.
When detecting, the received target signals at the radar are buried in the strong returned signals reflected by the sea surface, referred to as sea clutters~\cite{SeaClutter_CFAR_TGRS2017,SeaClutter_Marine_M}.
It is known that better detection performance is achieved if prior knowledge of sea clutters' distribution can be acquired, since by this a proper detection threshold can be determined at the detector~\cite{SeaClutter_TSP3,SeaClutter_Li_JSAC2016,SeaClutter_Li2_JSAC2016}.
Therefore, an important question that arises is how to accurately model the distribution of sea clutters in the fluctuating sea state for small target detection.


By adopting various parametric models~\cite{SeaClutter_handbook_1989,SeaClutter_compound_1981,SeaClutter_Gamma_TGRS2017,SeaClutter_Weibull_TAES1976}, there have been extensive works attempting to characterize the distribution of sea clutters to obtain better detection performance.
In~\cite{SeaClutter_handbook_1989}, the authors utilized the Gauss distribution to model the amplitude of the sea clutter at a low-resolution radar. With an increase in radar's spatial resolution, the amplitude distribution of the sea clutters was further extended from the Gaussian to compound-Gaussian probability density functions (PDFs) in~\cite{SeaClutter_compound_1981} for small target detection.
Gao \textit{et al.} in~\cite{SeaClutter_Gamma_TGRS2017} adopted the generalized Gamma distribution to describe the statistical behaviors of sea clutters, and provided a parameter estimation scheme by taking both of estimation precision and applicable conditions into consideration.
In~\cite{SeaClutter_Weibull_TAES1976}, the authors adopted the Weibull model in the constant false alarm rate (CFAR) detector for radar detection and evaluated the involved parameter optimization problem.

To summarize, the distributions of sea clutters adopted in~\cite{SeaClutter_handbook_1989,SeaClutter_compound_1981,SeaClutter_Gamma_TGRS2017,SeaClutter_Weibull_TAES1976}
for target detection are generally established as parametric models.
However, considering the two following defects of the parametric models, a considerable gap possibly exists between the fitted and realistic distribution of sea clutters. Firstly, the parametric models can hardly depict the spiky components in sea clutters, which is produced when the high-resolution radar works at a low grazing angle or under dynamic sea states~\cite{SeaClutter_spiky_TAP2015}. Secondly, as the distribution of sea clutters usually varies with different detection environments, assuming a fixed parametric model for it cannot guarantee satisfactory fitting performance in the varying detection environments and thus degrades the detection performance.

Following these insights, the distribution of sea clutters should be characterized by sufficiently analyzing the collected data instead of assuming a parametric model.
Inspired by this, the kernel density estimation (KDE), a non-parametric approach for estimating the PDF of a random variable, can be adopted to reveal the distribution of the collected sea clutters. Different from the methods in~\cite{SeaClutter_handbook_1989,SeaClutter_compound_1981,SeaClutter_Gamma_TGRS2017,SeaClutter_Weibull_TAES1976}, the KDE method utilizes smooth kernel functions to fit the realistic distribution of the observed data without making any assumption on it, which can effectively reflect the information of the spiky components and flexibly adapt to the varying detection environments.


When applying the KDE method, it is of utmost importance to determine two key parameters, namely the kernel function and the bandwidth~\cite{SeaClutter_KDE_TKDE2017}.
In~\cite{SeaClutter_Plugin_JASA1996}, the Gaussian kernel and some traditional bandwidth selectors such as the plug-in were adopted in the KDE method, which show good fitting performance on random sequence samples.
However, few works have ever studied how the KDE method works in the sea-surface target detection.
In addition, whether there are other kernel functions that can achieve better fitting performance than the Gaussian kernel or not is still unclear.
Furthermore, it is also quite challenging to derive the optimal bandwidth for other specialized kernels by the traditionally complicated bandwidth selection methods such as the plug-in technique~\cite{SeaClutter_Plugin_JASA1996}. These challenges impose restrictions on the application of KDE method in estimating the distribution of sea clutters.

In view of these, this paper first develops a KDE-based sea clutter modeling framework that is suitable for different kernel functions. In this framework, two embedded fundamental problems, the selection of a proper kernel density function and the determination of its corresponding optimal bandwidth, are needed to be solved.
Considering three kinds of kernels, i.e., Gaussian, Gamma, and Weibull, we then derive their respective closed-form optimal bandwidth equations and design a fast iterative bandwidth selection algorithm to solve them.

The main contributions of this work are as follows:
\begin{itemize}
 \item We propose a KDE-based framework that enfolds kernel function selection and bandwidth optimization to precisely model the sea clutter distribution. Compared with traditional parametric methods, this framework can not only take the information of spiky components into account but also adapt to varying detection environments.

 \item Inspired by parametric sea clutter models, we select the Gaussian, Gamma, and Weibull distributions as the kernels in our proposed framework. Particularly, we derive closed-form equations of the optimal bandwidth for these three kernels, which are unlikely to be deduced by adopting traditional bandwidth selection methods such as the plug-in technique. Due to the high complexity in solving these derived equations, we further design a fast iterative bandwidth selection algorithm to calculate the optimal bandwidth for each of kernels.

 \item Experimental results exhibit that our proposed approach outperforms the existing methods in terms of the modeling error (about two orders of magnitude reduction). Moreover, applying our modeled sea clutter distribution into the CFAR detector can significantly improve the detection probability, especially in low false alarm rate cases (up to 36\%).

\end{itemize}

\section{System Scenarios and Problem Formulation} \label{Section:SystemModel}
In this section, we first introduce the realistic Intelligent PIxel processing X-band (IPIX) radar datasets, and then formulate the asymptotic mean integrated square error (AMISE) minimization problem.

\subsection{IPIX Datasets} \label{Subsection:IPIX}
In this paper, we adopt the IPIX database, an authoritative and widely-used database collected at the east coast of Canada in November 1993, to model the distribution of sea clutters. As shown from a website held by Simon Haykin~\cite{IPIX_website_1993}, there are a total of 14 datasets in the collected database. Each dataset includes 14 separate spatial range cells, with each cell containing a length of 131072 time sampling data. These cells can be divided into three categories. Specifically, the cell with the target signals is labeled as the primary cell, the adjacent cells affected by the target are labeled as the secondary cells, and the remaining cells are clutter-only cells. For notational simplicity, we denote the samples from the primary cells and clutter-only cells as target signals and sea clutters, respectively.

\subsection{Statistical Sea Clutter Distributions} \label{Subsection:The PDFs of Statistical Distribution}
As is known to all, the amplitude of sea clutters usually follows Gaussian distribution for a low-resolution radar. However, it was soon found that the Gaussian distribution exhibits a poor fitting performance as the spatial resolution of the radar increases. Moreover, the amplitude distribution of sea clutters at a high-resolution radar
will demonstrate the characteristics of the compound-Gaussian models. These models, e.g., the Gamma and Weibull models, have been widely used for sea clutter distribution modeling and shown a better fitting performance compared with the Gaussian distribution. In what follows, we present the PDFs and fitting performance of the above distributions.

\subsubsection{Gaussian distribution}
The PDF of the Gaussian distribution is given as
\begin{equation} \label{Eq:GaussianModel}
f\left( {x|\mu ,{\sigma ^2}} \right) = \frac{1}{{\sqrt {2\pi {\sigma ^2}} }}{e^{ - \frac{{{{\left( {x - \mu } \right)}^2}}}{{2{\sigma ^2}}}}}
\end{equation}
where $\mu$ is the expectation of the distribution, $\sigma$ is the standard deviation, and $x$ denotes the amplitude of the clutter samples.

\subsubsection{Gamma distribution}
The PDF of the Gamma distribution is presented as
\begin{equation} \label{Eq:GammaModel}
f\left( {x|\alpha ,\beta } \right) = \frac{{{\beta ^\alpha }}}{{\Gamma \left( \alpha  \right)}}{x^{\alpha  - 1}}{e^{ - \beta x}}
\end{equation}
where $\alpha>0$ and $\beta>0$ are the shape parameter and the rate parameter of the Gamma distribution, respectively. In addition, $\Gamma \left( \alpha  \right)$ represents the complete gamma function.

\subsubsection{Weibull distribution}
The PDF of the Weibull distribution can be described as
\begin{equation} \label{Eq:WeibullModel}
f\left( {x|c ,s} \right) = \left\{ {\begin{array}{*{20}{c}}
{\frac{s}{c }{{\left( {\frac{x}{c }} \right)}^{s - 1}}{e^{ - {{\left( {x/c } \right)}^s}}}\quad x \ge 0}\\
{0\quad \quad \quad \quad \quad \quad \ \ \;\;\,\,x < 0}
\end{array}} \right.
\end{equation}
where $s>0$ and $c>0$ are the shape parameter and the scale parameter of the Weibull distribution, respectively.

\begin{figure}[t]
\centering \leavevmode \epsfxsize=3.5 in  \epsfbox{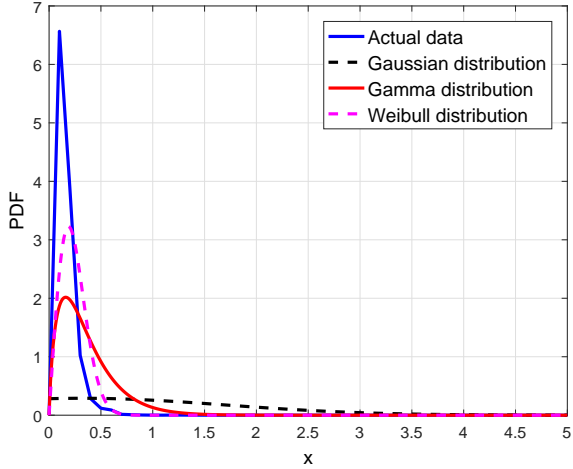}
\centering \caption{Comparison among the three PDFs produced by individually applying the Gaussian, Gamma, and  Weibull distributions to estimate the actual PDF of the IPIX data, where $\sigma=1.38$ and $\mu=0.32$ are selected for the Gaussian distribution, $\alpha=1.8$ and $\beta=0.2$ for the Gamma distribution, and $c=0.27$ and $s=2$ for the Weibull distribution.} \label{Fig:traditional_pdf}
\end{figure}

To obtain the best combination of parameters for the aforementioned three functions, we first utilize~(\ref{Eq:GaussianModel}), (\ref{Eq:GammaModel}), and (\ref{Eq:WeibullModel}) to model the distribution of sea clutter data based on the mean squared error (MSE) criterion.
Under the best parameter settings, we then calculate their corresponding PDFs, which are plotted in Fig.~\ref{Fig:traditional_pdf}.
From the figure, the fitting performance of the Gamma and Weibull distributions are better than the Gaussian distribution in the case when the normalized amplitude is more than 0.
However, there still exists considerable bias between the curves of statistical PDFs and the practical sea clutter PDF, especially in the cases of low normalized amplitude.
This phenomenon implies that the distribution of sea clutters should be estimated by tracking and characterizing the instant changes instead of traditionally assuming a parametric model.

\subsection{Problem Formulation} \label{Subsection:Problem_formulation}

We firstly utilize the kernel density estimation, a nonparametric approach for estimating the probability density function of a random variable, to model the distribution of sea clutters. By adopting the kernel function and bandwidth, the KDE approach assigns a height curve to each observation point. Each curve needs to be normalized first and then summed up by the kernel estimator function to estimate the density of sea clutters.
The expression of KDE can be written as follows~\cite{SeaClutter_KDE_TKDE2017}
\begin{equation} \label{Eq:KDE_Model}
\hat f\left( x \right) = \frac{1}{{Nh}}\sum\limits_{i = 1}^N {K\left( {\frac{{x - {x_i}}}{h}} \right)}
\end{equation}
where $K$ is the kernel density function, $h$ is the bandwidth of the KDE method, $N$ is the number of sample points, and $x_i$ is the $i$-th sample point.

Generally, the kernel density function and the bandwidth are two key factors that determine the estimation performance. For a given kernel density function, there exists an optimal bandwidth that achieves the best estimation accuracy, and larger or smaller bandwidth will lead to worse fitting performance. In what follows, we denote $f\left( x \right)$ as the density function of sea clutters and demonstrate the procedure of the theoretical derivation of the optimal bandwidth.

To derive the expression of optimal bandwidth, we then introduce a useful criterion, referred to as the AMISE, to evaluate the fitting performance of the distribution estimation $\hat f\left( x \right)$, given by~\cite{SeaClutter_AMISE_1986}
\begin{equation} \label{Eq:AMISE}
\text{AMISE}\left\{ {\hat f\left( x \right)} \right\} = \frac{1}{{Nh}}R\left( K \right) + \frac{1}{4}{h^4}{\mu _2}{\left( K \right)^2}R\left( {f''} \right)
\end{equation}
where $R\left( K \right) = \int_R {K{{\left( x \right)}^2}dx} $, ${\mu _2}\left( K \right) = \int_R {{x^2}K\left( x \right)dx} $, and $f''$ is the second derivative of the density $f$.
Then the optimal bandwidth can be straightforward derived by an optimization problem as follows
\begin{equation} \label{Eq:h_AMISE}
{h_\text{AMISE}} = \arg \mathop {\min }\limits_h \text{AMISE}\left\{ {\hat f\left( x \right)} \right\}.
\end{equation}
The solution of~(\ref{Eq:h_AMISE}) is illustrated in the following theorem.

\begin{theorem} \label{theorem1}
For each kernel density function to estimate the unknown density $f$, there exists a general expression for the optimal bandwidth, given by
\begin{equation} \label{Eq:opt_bandwidth}
{h_{\rm{opt}}} = {\left[ {\frac{{R\left( K \right)}}{{{\mu _2}{{\left( K \right)}^2}R\left( {f''} \right)N}}} \right]^{1/5}}.
\end{equation}
\end{theorem}

\begin{IEEEproof}
Proof of this theorem can be found in~\cite{SeaClutter_AMISE_1986}.
\end{IEEEproof}


Theorem~\ref{theorem1} indicates that $R\left( K \right)$, ${\mu _2}\left( K \right)$, and $R\left( {f''} \right)$ should be precalculated when determining the optimal bandwidth of the KDE method.
Among these variables, although the $R\left( K \right)$ and ${\mu _2}\left( K \right)$ can be easily derived if the kernel function is given, the value of $R\left( {f''} \right)$ is difficult to obtain as the density $f$ is still unknown to us.
To solve this problem, the estimation of $R\left( {f''} \right)$ is used to transform $R\left( {f''} \right)$ into an easy-to-calculate expression, which will be illustrated in the next section.

\section{Derivation of Optimal Bandwidths for Different Kernels} \label{Section:Derivation}
In this section, we propose an analytical approach to solve problem~(\ref{Eq:h_AMISE}) in Section~\ref{Subsection:Problem_formulation}, where the derived optimal bandwidth will be varied with different kinds of kernels. Interestingly, statistical models, e.g., Gaussian, Gamma, and Weibull distributions, usually can reveal the physical nature of sea clutters~\cite{SeaClutter_Iteration_TAES2017}. Inspired by this, we take the Gaussian, Gamma, and Weibull kernels as examples to evaluate the fitting performance of the KDE-based method.

\subsection{Gaussian Kernel Density Function}
According to the Gaussian distribution~(\ref{Eq:GaussianModel}), the kernel density function $K$ can be expressed as
\begin{equation} \label{Eq:Gaussian_distribution}
K\left( {x} \right) = \frac{1}{{\sqrt {2\pi {\sigma ^2}} }}{e^{ - \frac{{{{\left( {x - \mu } \right)}^2}}}{{2{\sigma ^2}}}}}.
\end{equation}

In order to derive the optimal bandwidth $h_\text{opt}$ in~(\ref{Eq:opt_bandwidth}), $R\left( K \right)$ and ${\mu _2}\left( K \right)$ should be derived in the first place, which will be quantified in the following lemma.

\begin{lemma}\label{lemma_Gaussian}
For the Gaussian kernel, $R\left( K \right)$ and ${\mu _2}\left( K \right)$ can be expressed as
\begin{equation} \label{Eq:Gaussian_R(K)}
R\left( K \right) = \int_{ - \infty }^\infty  {{K^2}\left( x \right)} dx = \frac{1}{{2\sqrt \pi  \sigma }}
\end{equation}
\begin{equation} \label{Eq:Gaussian_u(K)}
{\mu _2}\left( K \right) = \int_{ - \infty }^\infty  {{x^2}K\left( x \right)} dx = {\sigma ^2} + {\mu ^2}.
\end{equation}
\end{lemma}

Compared with $R\left( K \right)$ and ${\mu _2}\left( K \right)$, it is more difficult to calculate $R\left( {f''} \right)$ due to the lack of the prior knowledge of the unknown density $f$. Our idea is to first reshape $R\left( {f''} \right)$ in an easy-to-calculate form and then estimate it via the estimator of the second derivative of the density.
The following theorem quantifies the optimal bandwidth for the Gaussian kernel.

\begin{theorem}\label{theorem2}
The \text{AMISE} gets its minimum for the Gaussian kernel density function when ${h_{\rm{opt}}} =  {\frac{{{{\left( {{\sigma ^2} + {\mu ^2}} \right)}^2}Q\left( h_{\rm{opt}}\right)}}{{\sqrt \pi {\sigma ^5}N}}}$, where $Q\left( h_{\rm{opt}}\right)$ is a function of $h_{\rm{opt}}$.
\end{theorem}

\begin{IEEEproof}
As concluded above, the AMISE gets its minimum when the optimal bandwidth $h_{\rm{opt}}$ is adopted in~(\ref{Eq:KDE_Model}). Substituting $h_{\rm{opt}}$ into $\hat f\left( x \right)$, the optimal estimation distribution $\hat f_{\rm{opt}}\left( x \right)$ of an unknown density $f$ is given by
\begin{equation} \label{Eq:f_opt}
\hat f_{\rm{opt}}\left( x \right) = \frac{1}{{Nh_{\rm{opt}}}}\sum\limits_{i = 1}^N {K\left( {\frac{{x - {x_i}}}{h_{\rm{opt}}}} \right)}.
\end{equation}
Based on~(\ref{Eq:f_opt}), the $m$-order derivatives of $\hat f_{\rm{opt}}\left( x \right)$ with respect to $x$ is calculated as~\cite{SeaClutter_Esti_2008}
\begin{equation} \label{Eq:KDE_mth}
{\hat f_{\rm{opt}}^{\left( m \right)}}\left( x \right) = \frac{1}{{N{h_{\rm{opt}}^{m + 1}}}}\sum\limits_{i = 1}^N {{K^{\left( m \right)}}\left( {\frac{{x - {x_i}}}{h_{\rm{opt}}}} \right)}
\end{equation}
where $K^{\left( m \right)}$ is the $m$-th derivative of the kernel $K$.

To transform the integral of the squared second derivative of $f$, i.e., $R\left( {f''} \right)$, into an easy-to-calculate expression, we utilize the function ${\hat f_{\rm{opt}}^{(m)}}\left( x \right)$ in~(\ref{Eq:KDE_mth}) to estimate it, given by
\setlength{\abovedisplayskip}{0pt}
\begin{spacing}{1.2}
\begin{equation} \label{Eq:R(f)_estimate}
\begin{small}
\begin{aligned}
R\left( {f''} \right) &\approx {\int {\left( {\hat f_{\rm{opt}}^{\left( 2 \right)}\left( x \right)} \right)} ^2}dx\\
 &= \frac{1}{{2\pi {{\left( {N{h_{\rm{opt}}^3}{\sigma ^3}} \right)}^2}}} \cdot \\
 &\ \ \ {\!\int\! {\left( {\sum\limits_{i = 1}^N {\left( {\frac{{{{\left( {{z_i} - \mu } \right)}^2}}}{{{\sigma ^2}}} - 1} \right)\!\!\exp\!\! \left( { - \frac{{{{\left( {{z_i} - \mu } \right)}^2}}}{{2{\sigma ^2}}}} \right)} } \right)} ^2}\!\!\!dx
\end{aligned}
\end{small}
\end{equation}
\end{spacing}
\noindent where
\begin{equation} \label{Eq:zi}
{z_i} = \frac{{x - {x_i}}}{h_{\rm{opt}}}.
\end{equation}
For notational simplicity, we set
\begin{equation} \label{Eq:pi}
{P_i} = \left( {\frac{{{{\left( {{z_i} - \mu } \right)}^2}}}{{{\sigma ^2}}} - 1} \right)\exp \left( { - \frac{{{{\left( {{z_i} - \mu } \right)}^2}}}{{2{\sigma ^2}}}} \right).
\end{equation}
Rearranging $R(f'')$ in~(\ref{Eq:R(f)_estimate}) yields
\begin{equation} \label{Eq:R(f)_P}
R\left( {f''} \right) = \frac{1}{{2\pi {{\left( {N{h_{\rm{opt}}^3}{\sigma ^3}} \right)}^2}}}{\int {\left( {\sum\limits_{i = 1}^N {{P_i}} } \right)} ^2}dx.
\end{equation}
Substituting the variable $z_i$~(\ref{Eq:zi}) and $P_i$~(\ref{Eq:pi}) into~(\ref{Eq:R(f)_P}), it is clearly seen that~(\ref{Eq:R(f)_P}) is related to $h_{\rm{opt}}$ if the parameters $\sigma$, $\mu$, and $N$ are fixed.

Let
\begin{equation} \label{Eq:Gaussian_Q}
Q\left( h_{\rm{opt}}\right) = {\int {\left( {\sum\limits_{i = 1}^N {{P_i}} } \right)} ^2}dx.
\end{equation}
Substituting~(\ref{Eq:Gaussian_R(K)}), (\ref{Eq:Gaussian_u(K)}), (\ref{Eq:R(f)_P}), and~(\ref{Eq:Gaussian_Q}) into~(\ref{Eq:opt_bandwidth}), we obtain the optimal bandwidth $h_\text{opt}$ based on Lemma~\ref{lemma_Gaussian}, given by
\begin{equation} \label{Eq:OPT_fin_h}
{h_{\rm{opt}}} =  {\frac{{{{\left( {{\sigma ^2} + {\mu ^2}} \right)}^2}Q\left( h_{\rm{opt}}\right)}}{{\sqrt \pi {\sigma ^5}N}}}.
\end{equation}

\end{IEEEproof}

\subsection{Gamma Kernel Density Function}

For the case of Gamma distribution, we first deduce the $R\left( K \right)$ and ${\mu _2}\left( K \right)$ in the following lemma.

\begin{lemma}\label{lemma_Gamma}
For the Gamma kernel, $R\left( K \right)$ and ${\mu _2}\left( K \right)$ can be expressed as
\begin{equation} \label{Eq:gamma_R(k)}
R\left( K \right) = {\left( {\frac{{{\beta ^\alpha }}}{{\Gamma \left( \alpha  \right)}}} \right)^2}\frac{{\Gamma \left( {2\alpha  - 1} \right)}}{{{{\left( {2\beta } \right)}^{2\alpha  - 1}}}}
\end{equation}
\begin{equation} \label{Eq:gamma_u(k)}
{\mu _2}\left( K \right) = \frac{{\Gamma \left( {\alpha {\rm{ + 2}}} \right)}}{{\Gamma \left( \alpha  \right)}}.
\end{equation}
\end{lemma}

Then, we derive the $h_{\rm{opt}}$ for the Gamma kernel based on Lemma~\ref{lemma_Gamma}. Similar to the proof of Theorem~\ref{theorem2}, we first utilize~(\ref{Eq:KDE_mth}) to estimate $R\left( {f''} \right)$ for the Gamma kernel, and then obtain $W\left( h_{\rm{opt}}\right)$ and $h_{\rm{opt}}$ (corresponding to $Q(h_{\rm{opt}})$ and $h_{\rm{opt}}$ in~(\ref{Eq:Gaussian_Q}) and~(\ref{Eq:OPT_fin_h}), respectively).
Related results are summarized in the following theorem and we omit its proof for brevity.

\begin{theorem} \label{theorem3}
The \text{AMISE} gets its minimum for the Gamma kernel density function when ${h_{\rm{opt}}} =  {\frac{{{{\left( {2\beta } \right)}^{2\alpha  - 1}}{\Gamma ^{\rm{2}}}\left( {\alpha {\rm{ + 2}}} \right)W\left( h_{\rm{opt}}\right)}}{{\Gamma \left( {{\rm{2}}\alpha {\rm{ - 1}}} \right){\Gamma ^{\rm{2}}}\left( \alpha  \right)N}}}$, where $W\left( h_{\rm{opt}}\right) = {\int {\left( {\sum\nolimits_{i = 1}^N {{G_i}} } \right)} ^2}dx$, ${z_i} = \frac{{x - {x_i}}}{h_{\rm{opt}}}$, and ${G_i} = {e^{ - \beta {z_i}}} \bigg\{ \left( {\alpha  - 1} \right)\left( {\alpha  - 2} \right)z_i^{\alpha  - 3} - 2\beta \left( {\alpha  - 1} \right)z_i^{\alpha  - 2} + {\beta ^2}z_i^{\alpha  - 1} \bigg\}$.
\end{theorem}

\subsection{Weibull Kernel Density Function}

As for Weibull kernel density function, we first deduce the $R\left( K \right)$ and ${\mu _2}\left( K \right)$ in the following lemma.

\begin{lemma} \label{lemma_Weibull}
For the Weibull kernel, $R\left( K \right)$ and ${\mu _2}\left( K \right)$ can be expressed as
\begin{equation} \label{Eq:weibull_R(K)}
R\left( K \right) = \frac{{\Gamma \left( {2 - \frac{1}{s}} \right)s}}{{{2^{2 - \frac{1}{s}}}c}}
\end{equation}
\begin{equation} \label{Eq:weibull_u(K)}
{\mu _2}\left( K \right) = {c^2}\Gamma \left( {\frac{2}{s} + 1} \right).
\end{equation}
\end{lemma}

Based on Lemma~\ref{lemma_Weibull}, we then quantify the optimal bandwidth for the Weibull kernel in the following theorem, the proof of which is similar to those of Theorem~\ref{theorem2} and~\ref{theorem3}.

\begin{theorem} \label{theorem4}
The \text{AMISE} gets its minimum for the Weibull kernel density function when ${h_{\rm{opt}}} =  {\frac{{{2^{2 - \frac{1}{s}}}{\Gamma ^2}\left( {\frac{2}{s} + 1} \right)sV\left( h_{\rm{opt}}\right)}}{{\Gamma \left( {2 - \frac{1}{s}} \right)cN}}}$, where $V\left( h_{\rm{opt}}\right) = {\int {\left( {\sum\nolimits_{i = 1}^N {{L_i}} } \right)} ^2}dx$, ${L_i} = {e^{ - {{\left( {\frac{{{z_i}}}{c}} \right)}^s}}}{\left( {\frac{{{z_i}}}{c}} \right)^{s - 3}} \cdot \bigg\{ {\left( {s - 1} \right)\left( {s - 2} \right) - 3s\left( {s - 1} \right){{\left( {\frac{{{z_i}}}{c}} \right)}^s} + {s^2}{{\left( {\frac{{{z_i}}}{c}} \right)}^{2s}}} \bigg\}$, ${z_i} = \frac{{x - {x_i}}}{h_{\rm{opt}}}$.
\end{theorem}

\begin{table}[t] \label{table1}
\footnotesize
\centering
\caption{Optimal bandwidth for different kernel density functions.}\label{Table:FittingResults1}
\begin{supertabular}{|c|c|c|}
\specialrule{0.05em}{3pt}{0pt}
Kernel functions & Optimal bandwidth \\
\specialrule{0.05em}{0pt}{0pt}
Gaussian  & ${h_{\rm{opt}}} =  {\frac{{{{\left( {{\sigma ^2} + {\mu ^2}} \right)}^2}Q\left( h_{\rm{opt}}\right)}}{{\sqrt \pi {\sigma ^5}N}}} = {\psi _1}\left( {h_{\rm{opt}}} \right)$  \\
\specialrule{0.05em}{0pt}{0pt}
Gamma  & ${h_{\rm{opt}}} =  {\frac{{{{\left( {2\beta } \right)}^{2\alpha  - 1}}{\Gamma ^{\rm{2}}}\left( {\alpha {\rm{ + 2}}} \right)W\left( h_{\rm{opt}}\right)}}{{\Gamma \left( {{\rm{2}}\alpha {\rm{ - 1}}} \right){\Gamma ^{\rm{2}}}\left( \alpha  \right)N}}} = {\psi _2}\left( {h_{\rm{opt}}} \right)$  \\
\specialrule{0.05em}{0pt}{0pt}
Weibull  & ${h_{\rm{opt}}} =  {\frac{{{2^{2 - \frac{1}{s}}}{\Gamma ^2}\left( {\frac{2}{s} + 1} \right)sV\left( h_{\rm{opt}}\right)}}{{\Gamma \left( {2 - \frac{1}{s}} \right)cN}}} = {\psi _3}\left( {h_{\rm{opt}}} \right)$  \\
\specialrule{0.05em}{0pt}{3pt}
\end{supertabular}
\end{table}

\begin{remark}\label{Remark:Weibull_kernel}
Based on Theorems~\ref{theorem2}, \ref{theorem3}, and \ref{theorem4}, we have rigorously derived the closed-form equations about the optimal bandwidths for the three different kernels with their involved parameters in Table~\ref{Table:FittingResults1}.
It is observed from the table that, the optimal bandwidths are all determined once the parameters of the selected kernel density functions are given.
Moreover, the expressions of optimal bandwidths are all in the form of the fixed-point equations since the $Q\left( h_{\rm{opt}}\right)$, $W\left( h_{\rm{opt}}\right)$, and $V\left( h_{\rm{opt}}\right)$ are the functions of $h_{\rm{opt}}$. This phenomenon motivates us to apply the fixed-point iterative algorithm to obtain $h_{\rm{opt}}$.
\end{remark}

\subsection{Algorithms for the Optimal Bandwidth Selection}
\label{Subsection:optimal_bandwidth_derivation}
In this subsection, we calculate the optimal bandwidths for the three kernels based on equations in Table~\ref{Table:FittingResults1}.
However, it is difficult to obtain a straightforward expression for the optimal bandwidth as the $Q\left( h_{\rm{opt}}\right)$, $W\left( h_{\rm{opt}}\right)$, and $V\left( h_{\rm{opt}}\right)$ are the functions of $h_{\rm{opt}}$, which are complicated and unsolvable.
Furthermore, traditional numerical analysis such as the Newton-Raphson method that is based on the derivation of the target equation also increases the computation complexity and reduces the efficiency of the KDE.

In view of these, we adopt the fixed-point theory to efficiently solve the equations in Table~\ref{Table:FittingResults1}, where the optimal bandwidths for the three kernels can be obtained by deriving the fixed-points of the equations.
To determine these fixed-points efficiently, we further design a fast iterative bandwidth selection algorithm, which is described in Algorithm~\ref{Algorithm:BandwidthSelectionAlgorithm}.

\begin{algorithm}[!t]
\caption{Bandwidth selection algorithm.}
\begin{algorithmic}[1] \label{Algorithm:BandwidthSelectionAlgorithm}
\STATE \textbf{Initialization} \label{line1}
\begin{itemize}
\item Set the started approximation bandwidth ${h_{l,0}}$ and the tolerance TOL.
\item Set the maximum number of iteration $N_{\max}$ and the iteration index $i=1$.
\end{itemize}
\STATE Denote ${\psi _1}$, ${\psi _2}$, and ${\psi _3}$ as the optimal bandwidth functions for the Gaussian, Gamma, and Weibull kernels, which are referred to in Table~\ref{Table:FittingResults1}. \label{line2}
\STATE \textbf{while} $(i \le {N_{\max }})$ \textbf{do} \label{line3}
\STATE \quad Set $h_{l} = {\psi _l}\left( {h_{l,0}} \right)$, where $l=$1, 2, or 3. \label{line4}
\STATE \quad \textbf{if} $\left| {\frac{{h_{l} - h_{l,0}}}{h_{l}}} \right| < \text{TOL}$ \textbf{then} \label{line5}
\STATE \quad \quad Set $H_l=h_{l,0}$. \label{line6}
\STATE \quad \quad break. \label{line7}
\STATE \quad \textbf{else} \label{line8}
\STATE \quad \quad Set $i = i + 1$ and ${h_{l,0}} = h_{l}$. \label{line9}
\STATE \quad \textbf{end if} \label{line10}
\STATE \textbf{end while} \label{line11}
\STATE Output the optimal bandwidth $H_l$. \label{line12}
\end{algorithmic}
\end{algorithm}


\section{Experimental Results and Analysis} \label{Section:SimulationResults}
In this section, we present experimental results to exhibit the fitting and detection performance of our proposed method using the realistic IPIX radar datasets.

Consider that the background sea clutter samples are needed to train the KDE model in Eq.~(\ref{Eq:KDE_Model}), the clutter-only cells in IPIX radar dataset, e.g., the 14-th cell in dataset 17, are thus adopted to be served as training sets. Based on the training sets, we obtain the best combination of parameters of the three kernel density functions under the MSE criterion, which are as follows: $\sigma=1.38$, $\mu=0.32$, $\alpha=1.8$, $\beta=0.2$, $c=0.27$, $s=2$, and $N=2048$.
Then, we utilize Algorithm~\ref{Algorithm:BandwidthSelectionAlgorithm} to search for the optimal bandwidths for different kernel functions.
The experimental results show that our proposed algorithm converges very fast (within 5 iterations) and the optimal bandwidths are equal to 0.08, 0.05, and 0.06 for the Gaussian, Gamma, and Weibull kernel functions, respectively.

\begin{figure}[t]
\centering \leavevmode \epsfxsize=3.5 in  \epsfbox{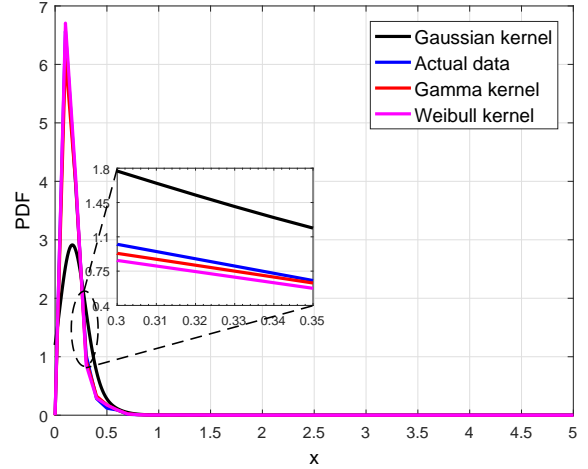}
\centering \caption{
Comparison among the three PDFs produced by individually applying the Gaussian, Gamma, and Weibull kernels to estimate the actual PDF of the IPIX data, where $\sigma=1.38$, $\mu=0.32$, and $h_{\text{opt}}=0.08$ are selected for the Gaussian kernel, $\alpha=1.8$, $\beta=0.2$, and $h_{\text{opt}}=0.05$ for the Gamma kernel, and $c=0.27$, $s=2$, and $h_{\text{opt}}=0.06$ for the Weibull kernel.} \label{Fig:KDE_result1}
\end{figure}

Based on the derived optimal bandwidths, we firstly compare the PDF fitting performance of our proposed KDE methods under different kernels in Fig.~\ref{Fig:KDE_result1}.
From the figure, it is obtained that, compared with the Gaussian kernel, the curves of Gamma and Weibull kernels are much closer to the distribution of sea clutters.
This phenomenon indicates that, if the optimal bandwidths are adopted for the three kernels, the kernel density function itself will play the dominant role in estimation.
Thus, it may be a better choice to select the Weibull and Gamma distributions rather than the Gaussian distribution as kernel functions.

\begin{figure}[t]
\centering \leavevmode \epsfxsize=3.5 in  \epsfbox{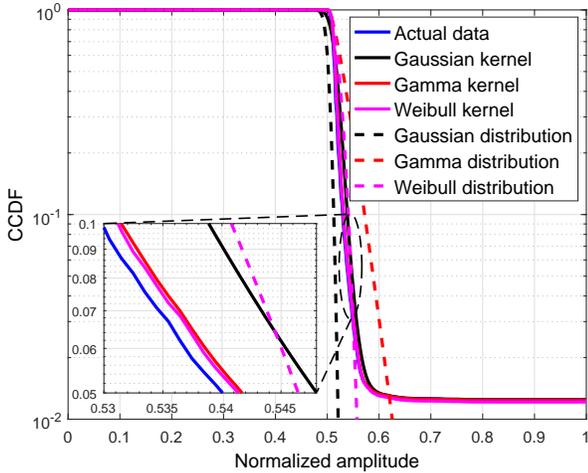}
\centering \caption{
Comparison among the six CCDFs produced by individually applying
the Gaussian, Gamma, and Weibull kernels and the Gaussian, Gamma, and Weibull distributions to estimate the actual CCDF of the IPIX data,
where the experimental parameters are the same as those in Fig.~\ref{Fig:KDE_result1}.} \label{Fig:KDE_result2}
\end{figure}

Secondly, we compare the complementary cumulative distribution function (CCDF) fitting performance of our proposed KDE methods for different kernels.
As depicted in Fig.~\ref{Fig:KDE_result2}, the CCDF curves of our proposed KDE methods are much closer to the sea clutter than those of the traditional parametric models, especially in high normalized amplitude.
Furthermore, the curves of the Gamma kernel and Weibull kernel almost overlap the curve of sea clutter, except for the Gaussian kernel which shows more or less difference from the others.
In particular, the obtained data shows that, compared with traditional parametric models, the Gaussian, Gamma, and Weibull kernels can reduce the MSE from the magnitude of $10^{-2}$ to the magnitudes of $10^{-3}$, $10^{-4}$, and $10^{-4}$, respectively.
Therefore, the Gamma and Weibull kernels are more preferred when applying the KDE methods to detect sea-surface targets.

Finally, we introduce the CFAR detector to test the detection performance $(P_{\text{d}})$ of our proposed KDE method. The CFAR detector works in a two-step process, namely the training and testing steps. In the training step, estimate the distribution of the clutter-only cell data and calculate the background level of sea clutters. Then, determine the threshold at a given false alarm rate $(P_{\text{fa}})$ by multiplying the background level with a given factor $T$. In the testing step, compare the amplitude of the testing samples with the obtained threshold to decide whether they are targets or not. We refer the readers to~\cite{SeaClutter_Iteration_TAES2017} on mechanisms for the detection and we omit it for brevity.

\begin{figure}[t]
\centering \leavevmode \epsfxsize=3.5 in  \epsfbox{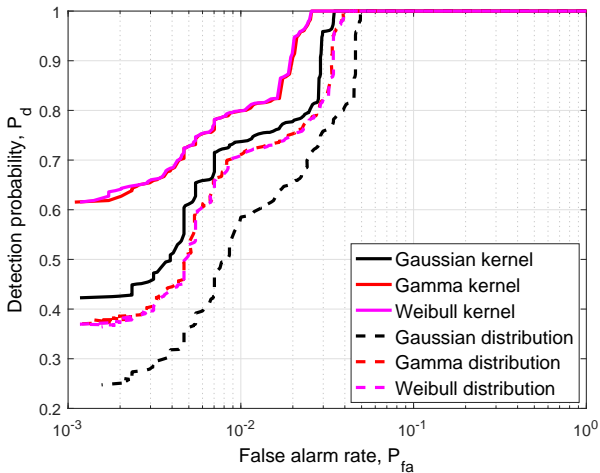}
\centering \caption{Comparison of the detection probability between our proposed method and traditional parametric modeling methods. In this figure, the testing samples are collected from dataset 17, and the CFAR detector is adopted for target detection.
} \label{Fig:KDE_detectionresults}
\end{figure}

Fig.~\ref{Fig:KDE_detectionresults} depicts how the detection probability of our proposed method and traditional parametric methods varies with the false alarm rate. From the figure, it is obtained that although the detection probabilities of these methods all increase with the false alarm rate, our proposed KDE method always achieves better detection performance than the others either in high or low false alarm rate cases. For example, our proposed Weibull kernel KDE method improves the $P_{\text{d}}$ by $23\%$, $24\%$, and $36\%$ compared with the Weibull, Gamma, and Gaussian distribution based methods, respectively, when the $P_{\text{fa}}$ is 0.001.


\section{Conclusions} \label{Section:Conclusions}
In this paper, we have put forward a KDE-based sea clutter modeling framework that is suitable for different kernel functions.
In this framework, we have firstly derived the closed-form optimal bandwidth equations for the Gaussian, Gamma, and Weibull kernels and then designed a fast iterative bandwidth selection algorithm to solve them. Experimental results have exhibited that, compared with existing methods, our proposed approach can significantly decrease the error incurred by sea clutter modeling (about two orders of magnitude reduction) and improve the target detection probability (up to $36\%$ in low false alarm rate cases).

\appendices

\bibliographystyle{IEEEtran}
\bibliography{IEEEabrv,1}

\begin{thebibliography}{10}
\providecommand{\url}[1]{#1}
\csname url@samestyle\endcsname
\providecommand{\newblock}{\relax}
\providecommand{\bibinfo}[2]{#2}
\providecommand{\BIBentrySTDinterwordspacing}{\spaceskip=0pt\relax}
\providecommand{\BIBentryALTinterwordstretchfactor}{4}
\providecommand{\BIBentryALTinterwordspacing}{\spaceskip=\fontdimen2\font plus
\BIBentryALTinterwordstretchfactor\fontdimen3\font minus
  \fontdimen4\font\relax}
\providecommand{\BIBforeignlanguage}[2]{{%
\expandafter\ifx\csname l@#1\endcsname\relax
\typeout{** WARNING: IEEEtran.bst: No hyphenation pattern has been}%
\typeout{** loaded for the language `#1'. Using the pattern for}%
\typeout{** the default language instead.}%
\else
\language=\csname l@#1\endcsname
\fi
#2}}
\providecommand{\BIBdecl}{\relax}
\BIBdecl

\bibitem{SeaClutter_Iteration_TAES2017}
W.~Zhou, J.~Xie, G.~Li, and Y.~Du, ``Robust {CFAR} detector with weighted
  amplitude iteration in nonhomogeneous sea clutter,'' \emph{IEEE Trans. Aero.
  Elec. Sys.}, vol.~53, no.~3, pp. 1520--1535, Jun. 2017.

\bibitem{SeaClutter_CFAR_TGRS2017}
G.~Gao and G.~Shi, ``{CFAR} ship detection in nonhomogeneous sea clutter using
  polarimetric {SAR} data based on the notch filter,'' \emph{IEEE Trans.
  Geosci. Remote}, vol.~55, no.~8, pp. 4811--4824, Aug. 2017.

\bibitem{SeaClutter_Marine_M}
Y.~Li, Y.~Zhang, W.~Li, and T.~Jiang, ``Marine wireless {Big} {Data}: Efficient
  transmission, related applications, and challenges,'' \emph{IEEE Wireless
  Commun.}, vol.~25, no.~1, pp. 19--25, Feb. 2018.

\bibitem{SeaClutter_TSP3}
S.~Haykin and T.~K. Bhattacharya, ``Modular learning strategy for signal
  detection in a nonstationary environment,'' \emph{IEEE Trans. Signal
  Process.}, vol.~45, no.~6, pp. 1619--1637, Jun. 1997.

\bibitem{SeaClutter_Li_JSAC2016}
Y.~Li, T.~Jiang, M.~Sheng, and Y.~Zhu, ``{QoS}-aware admission control and
  resource allocation in underlay device-to-device spectrum-sharing networks,''
  \emph{IEEE J. Sel. Areas Commun.}, vol.~34, no.~11, pp. 2874--2886, Nov.
  2016.

\bibitem{SeaClutter_Li2_JSAC2016}
Y.~Li, M.~Sheng, Y.~Sun, and Y.~Shi, ``Joint optimization of {BS} operation,
  user association, subcarrier assignment, and power allocation for
  energy-efficient {HetNets},'' \emph{IEEE J. Sel. Areas Commun.}, vol.~34,
  no.~12, pp. 3339--3353, Dec. 2016.

\bibitem{SeaClutter_handbook_1989}
F.~T. Ulaby and M.~C. Dobson, \emph{Handbook of Radar Scattering Statistics for
  Terrain.}\hskip 1em plus 0.5em minus 0.4em\relax Norwood, MA: Artech House,
  1989.

\bibitem{SeaClutter_compound_1981}
K.~Ward, ``Compound representation of high resolution sea clutter,''
  \emph{Electron. Lett.}, vol.~17, no.~16, pp. 561--563, Aug. 1981.

\bibitem{SeaClutter_Gamma_TGRS2017}
G.~Gao, K.~Ouyang, Y.~Luo, S.~Liang, and S.~Zhou, ``Scheme of parameter
  estimation for generalized {Gamma} distribution and its application to ship
  detection in {SAR} images,'' \emph{IEEE Trans. Geosci. Remote}, vol.~55,
  no.~3, pp. 1812--1832, Mar. 2017.

\bibitem{SeaClutter_Weibull_TAES1976}
D.~Schleher, ``Radar detection in {Weibull} clutter,'' \emph{IEEE Trans. Aero.
  Elec. Sys.}, vol. AES-12, no.~6, pp. 736--743, Nov. 1976.

\bibitem{SeaClutter_spiky_TAP2015}
Y.~Wei, L.~Guo, and J.~Li, ``Numerical simulation and analysis of the spiky sea
  clutter from the sea surface with breaking waves,'' \emph{IEEE Trans. Antenn.
  Propag.}, vol.~63, no.~11, pp. 4983--4994, Nov. 2015.

\bibitem{SeaClutter_KDE_TKDE2017}
A.~Qahtan, S.~Wang, and X.~Zhang, ``{KDE-Track}: An efficient dynamic density
  estimator for data streams,'' \emph{IEEE Trans. Knowl. Data En.}, vol.~29,
  no.~3, pp. 642--655, Mar. 2017.

\bibitem{SeaClutter_Plugin_JASA1996}
M.~C. Jones, ``A brief survey of bandwidth selection for density estimation,''
  \emph{J. Am. Stat. Assoc.}, vol.~91, no. 433, pp. 401--407, Jun. 1996.

\bibitem{IPIX_website_1993}
\BIBentryALTinterwordspacing
 [Online]. Available: \url{http://soma.ece.mcmaster.ca/ipix/.}
\BIBentrySTDinterwordspacing

\bibitem{SeaClutter_AMISE_1986}
B.~W. Silverman, \emph{Density estimation for statistics and data
  analysis}.\hskip 1em plus 0.5em minus 0.4em\relax Boca Raton, FL, USA:
  Chapman \& Hall/CRC, 1986.

\bibitem{SeaClutter_Esti_2008}
L.~A. Alexandre, ``A solve-the-equation approach for unidimensional data kernel
  bandwidth selection,'' Technical Report, University of Beira Interior,
  Portugal, 2008.

\end{thebibliography}
\end{document}